\documentclass{svproc}
\usepackage{url}

\usepackage{graphicx,epstopdf}
\usepackage{cite}
\usepackage{amssymb}
\usepackage{color}
\usepackage{float}
\usepackage{epsfig}
\usepackage{epstopdf}
\usepackage{framed}
\usepackage{subfigure}
\usepackage{enumerate}
\usepackage{empheq}

\usepackage[colorlinks=true, citecolor=blue, urlcolor=blue ]{hyperref}
\usepackage{cancel}
\def\beq{\begin{equation}}
\def\eeq{\end{equation}}\font\numbers=cmss12
\font\upright=cmu10 scaled\magstep1
\def\stroke{\vrule height8pt width0.4pt depth-0.1pt}
\def\topfleck{\vrule height8pt width0.5pt depth-5.9pt}
\def\botfleck{\vrule height2pt width0.5pt depth0.1pt}
\def\Zmath{\vcenter{\hbox{\numbers\rlap{\rlap{Z}\kern
0.8pt\topfleck}\kern 2.2pt
                   \rlap Z\kern 6pt\botfleck\kern 1pt}}}
\def\Qmath{\vcenter{\hbox{\upright\rlap{\rlap{Q}\kern
                   3.8pt\stroke}\phantom{Q}}}}
\def\Nmath{\vcenter{\hbox{\upright\rlap{I}\kern 1.7pt N}}}
\def\Cmath{\vcenter{\hbox{\upright\rlap{\rlap{C}\kern
                   3.8pt\stroke}\phantom{C}}}}
\def\Rmath{\vcenter{\hbox{\upright\rlap{I}\kern 1.7pt R}}}
\def\Hmath{\vcenter{\hbox{\upright\rlap{I}\kern 1.7pt H}}}
\def\Amath{\vcenter{\hbox{\upright\rlap{I}\kern 1.7pt A}}}
\def\Z{\ifmmode\Zmath\else$\Zmath$\fi}
\def\Q{\ifmmode\Qmath\else$\Qmath$\fi}
\def\N{\ifmmode\Nmath\else$\Nmath$\fi}
\def\C{\ifmmode\Cmath\else$\Cmath$\fi}
\def\R{\ifmmode\Rmath\else$\Rmath$\fi}

\def\middot{\textperiodcentered~}
\begin{document}
\mainmatter             

\title{Emergent geometry from  entanglement structure}

\author{Sudipto Singha Roy$^1$, Silvia N. Santalla$^2$, Javier Rodr{\'i}guez-Laguna$^3$, Germ{\'a}n Sierra$^{1,*}$}

\institute{$^1$Instituto de F{\'i}sica Te{\'o}rica, UAM-CSIC, Universidad Aut{\'o}noma de Madrid Cantoblanco, Madrid, Spain\\ $^2$ Dept. de F{\'i}sica and Grupo Interdisciplinar de Sistemas Complejos (GISC), Universidad Carlos III de Madrid, Spain \\$^3$ Dept. de  F{\'i}sica Fundamental, UNED, Madrid, Spain\\ \email{$^*$german.sierra@uam.es} }

\authorrunning{S. Singha Roy et al.}

\date{\today}

\maketitle              
\begin{abstract}
We attempt to reveal the geometry, emerged from the entanglement structure of any general $N$-party pure quantum many-body state by representing entanglement entropies corresponding to all $2^N $ bipartitions of the state by means of a generalized adjacency matrix. We show this representation is often exact and may lead to a geometry very different than suggested by the Hamiltonian. Moreover, in all the cases, it yields a natural entanglement contour, similar to previous proposals. The formalism is extended for conformal invariant systems, and a more insightful interpretation of entanglement is presented as a flow among different parts of the system.
\end{abstract}

\section{Introduction} 

Study of the  distribution of quantum entanglement in different parts of  the low-lying states of quantum many-body Hamiltonians often unveils many interesting features related to the physical system\cite{general_entang2,general_entang2a,general_entang2b}. For instance, bounded growth  of quantum entanglement between a region and its exterior  can be attributed to the fact that interactions in the quantum many-body systems are typically local\cite{area1,area2,area3,area4,our_paper1,our_paper2}. In this work, we aim to  explore the connection between the area-law for entanglement\cite{area1,area2,area3,area4} and  geometry which emerges from the distribution of quantum entanglement across all possible bipartitions of a pure quantum many-body state.  Towards this aim,  we first   define the notion of  geometry by means of a generalized adjacency matrix such that the approximate entanglement entropy of any given bipartition can be obtained as a linear sum of the weights of the links connecting it with its surroundings. We show that the representation is exact when there is a perfect area-law.  In other cases, it still provides an efficient approximation with minimal error. Interestingly, we also report some other important states, e.g., the {\it rainbow state}\cite{rainbow_bunch0,rainbow_bunch1,rainbow_bunch2,rainbow_bunch3,rainbow_bunch4,rainbow_bunch5,rainbow_bunch6,rainbow_bunch7}, where though a strong violation of area-law   is observed for the geometry defined by the local structure of the Hamiltonian, an area-law feature can indeed be recovered  for a geometry which is completely different than that suggested by the Hamiltonian. \

As an application of the formalism,  we provide a route to compute the entanglement contour function for quantum many-body systems, which is radically different than that previously introduced in Ref.\cite{contour1}. A quantitative comparison of the contour functions obtained using these two different approaches is made for the ground state of a non-interacting model. Additionally, we also study the behavior of contour function obtained for an interacting model, which surpasses the limitation of the previous approach\cite{contour1,contour2,contour3,contour4,contour5,contour6,contour7}.\

Finally, we extend our analysis to conformal invariant physical systems\cite{cft1,cft2,cft3,cft4}. As an important finding, we show that the conformal field theory  (CFT) descriptions help us to interpret the elements of the generalized adjacency matrix as the two-point correlator of an entanglement current operator. This field theory realization provides a  framework to consider entanglement as a flow among different parts of the system\cite{Cirac05}, similar to the flow of energy that is characterized by the stress tensor.\

In the following sections, after briefly introducing the formalism, we elaborate on our main findings.


\section{Emergent geometry} 
We start with an $N$-party pure quantum state $|\psi\rangle$, and characterize its entanglement properties by computing the von Neumann entropies $S_A=-\text{Tr}_A(\rho_A\log\rho_A)$ for all possible bipartitions of the state, namely $(A, A^c)$, where $\rho_A=\text{Tr}_{A^c}|\psi\rangle \langle \psi|$, and $\text{Tr}_{A (A^c)}$ denotes partial trace on the subsystem $A (A^c)$. We then aim to investigate whether the set of entropies obtained in this way respond to an area-law for some geometry. As a first step,  we assign a classical spin configuration $\{s_i\}^N$ to each such bipartitions using the rule\[s_i=\begin{cases}1,& \text{if } i \in A\\-1,   & \text{if }          i \in A^c.\end{cases}\]These spins are not physical but only a convenient way to describe the different bipartitions of the system. If two spins, say $i$ and $j$, belong to the same partition, $A$ or $A^c$, we get $s_is_j = 1$, while if they belong to different partitions, $s_is_j = -1$. In the former case, there is no contribution to the entanglement entropy $S_A$, while in the second case, they may contribute to $S_A$ with a certain amount that will depend on their positions. We are thus led to express the entropy $S_A$ of the bipartition $(A,A^c)$ as\begin{eqnarray}S_A= \frac{1}{2}\sum_{i  j }J_{ij}(1-s_is_j) + s_0,\label{eqn:Ising:classical}\end{eqnarray}where $J_{ij}$ defines the coupling between the classical spins $i$ and $j$ and $s_0$ may constitute a topological entropy term. The entropy $S_A$ thus can be further simplified  as the sum of contributions coming from all possible pairs, $J_{ij}$ i.e.,\begin{eqnarray}S_A=\sum_{i\in A, j\in A^c}J_{ij} + s_0.\label{entropy_general}\end{eqnarray}

\begin{figure}[h]
\begin{center}
  \includegraphics[width=5cm]{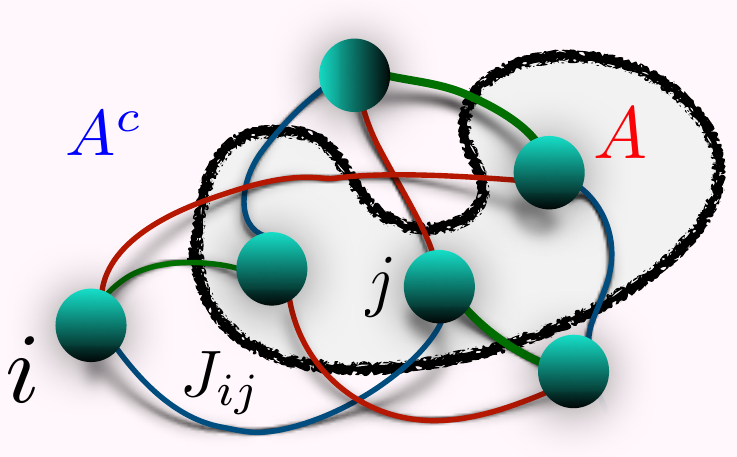}
\caption{Schematic of the entanglement entropy obtained for an arbitrary bipartition ($A, A^c$) by removing the links connecting the sites. Here the links represent the constants $J_{ij}$.}
  \label{schematic}
  \end{center}
  \end{figure}

A closer look at the derivation of the above entropy function reveals the fact that it is a clear manifestation of the area-law of entanglement entropy associated with the geometry revealed by the elements of $J$. More elaborately,   the coupling function $J_{ij}$ can be interpreted as the weight of an adjacency matrix of a generalized graph, such that the approximate entanglement entropy of the region $A$,  can be computed only by simply summing the weights ($J_{ij}$) associated with all the connecting edges between $A$ and $A^c$.  A schematic representation of the above formulation is depicted in Fig. \ref{schematic}. If Eq. (\ref{entropy_general}) holds exactly or at least approximately, the matrix $J$ will be termed the entanglement adjacency matrix (EAM) of the state $|\psi\rangle$. Additionally, we note that for the case when Eq. (\ref{entropy_general}) is exact, $J_{ij}$ equals to the mutual information between the sites $i, j$.
  

\subsection{Graphical representation} 
\begin{figure}[h]  
\begin{center}
  \includegraphics[width=8cm]{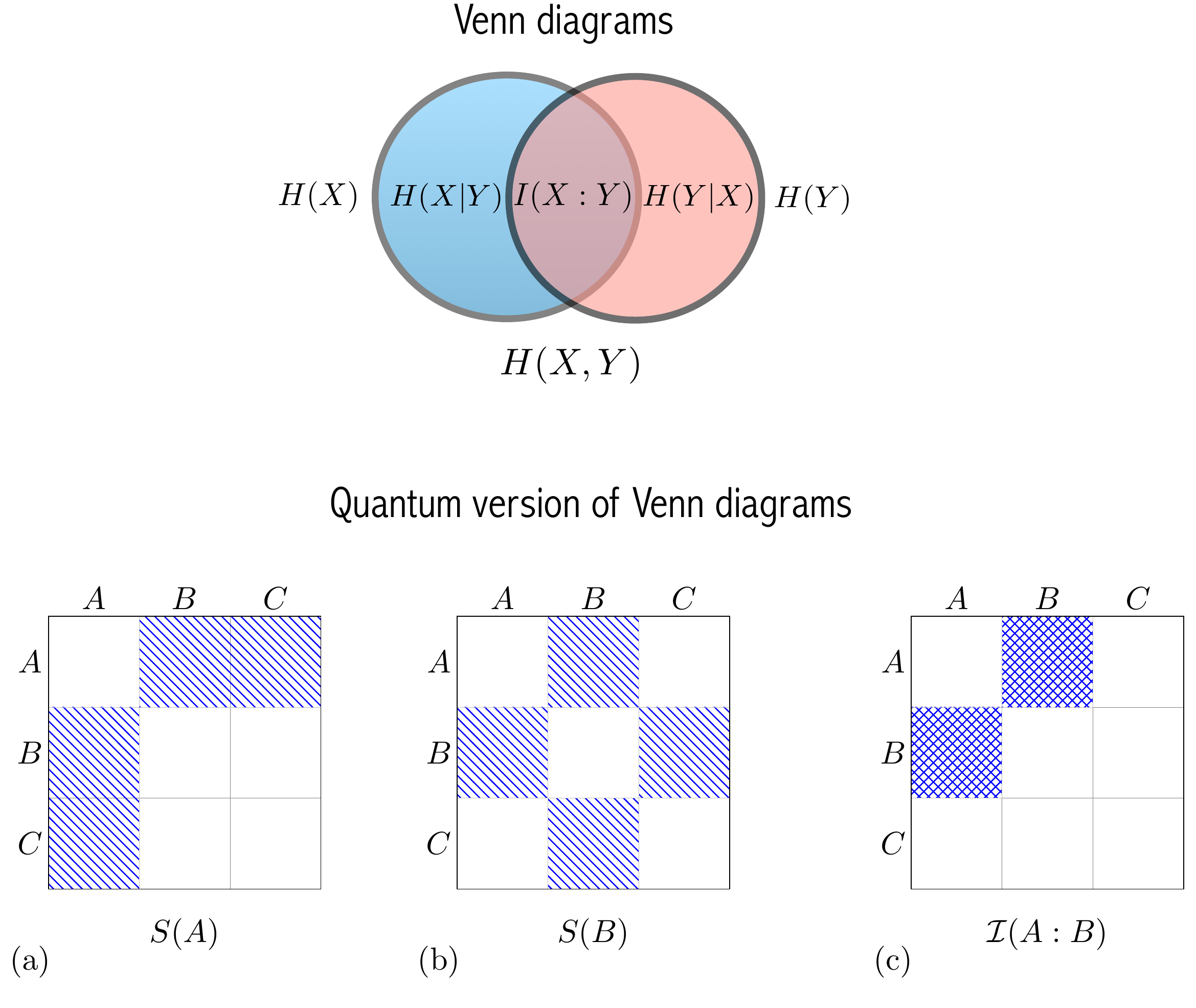}
  \caption{Similar to the illustration of the different entropies and the mutual information between two variables $X$ and $Y$ using Venn diagrams, as shown in the upper panel, a graphical representation of the quantum version of different entropic relations is presented in the lower panel (a)-(c).}
 \label{QVenn}
 \end{center}
  \end{figure}

An equivalent way to conceptualize the entropic functions obtained for different bipartitions is through the following graphical representation we outline here. Similar to Venn diagrams, which illustrate the logical relationships between two or more sets, here we present the schematic representation of the entropy values of different biapartitions by shading different regions in the $J$-matrix. As an example, consider a contiguous bipartition $(A,A^c)$, such that the sites $1 \dots m\in A$ and $m+1\dots N \in A^c$.  The entropy value of the block $A$ ($S_A$) can be schematically represented by shading the region of the $J$-matrix with $i \in 1,2,\dots m$ and $j\in m+1\in N$ and its conjugate part. Fig. \ref{QVenn} depicts the single-site entropies and the mutual information of a  system $AA^c$  by shading different regions in the $J$-matrix. 


\subsection{Exact examples} 

For any general pure quantum state, computation of its entanglement adjacency matrix requires knowledge of the entanglement of all possible bipartitions of the state. The number of such bipartitions increases exponentially with the size of the system. Hence,  even for moderate system size, estimation of the entanglement adjacency matrix demands lots of computational effort. However, if the quantum state possesses certain properties such that all entropies can be computed analytically, corresponding entanglement adjacency matrix can be obtained straightforwardly. Below we discuss few such cases, where the elements of the entanglement adjacency matrix can be computed exactly. \

{\it {\bf I.} Dimer model.-} We start with the {\it dimer state}, which can be mathematically expressed as $|\Psi\rangle=|i_1 j_1\rangle \otimes |i_2 j_2\rangle \otimes \dots \otimes |i_k j_k\rangle$, with $k=\frac{N}{2}$ for even $N$ and $|i_l j_l\rangle=\frac{1}{\sqrt{2}}(|\uparrow_{i_l} \downarrow_{j_l}\rangle-|\downarrow_{i_l} \uparrow_{j_l}\rangle)$.  Such states belong to the family of valence bond states and appear as approximate ground states of certain strongly inhomogeneous free-fermionic model.  Here, from the configuration of the state, one can  observe that all the single-site entropies become $S(\rho_i)=\log(2), \forall i \in N$. On the other hand, for the two-site blocks, if $i, j$ form a dimer $S(\rho_{ij})=0$, and $S(\rho_{ij})=2\log(2)$, otherwise. Therefore, the elements of the entanglement adjacency matrix possesses  non-zero values only when $i,j$ form singlet, given by$J_{ij}=\log(2)$. In this case, one can find that the geometry revealed by the entanglement adjacency matrix is a mere restriction of the one-dimensional adjacency matrix representing the Hamiltonian.\

{\it {\bf II.} Rainbow state.-} Another important member of the family of valence bond states we consider in our work  is the {\it rainbow state},    which is also the ground state of a local Hamiltonian\cite{rainbow_bunch0,rainbow_bunch1,rainbow_bunch2,rainbow_bunch3,rainbow_bunch4,rainbow_bunch5,rainbow_bunch6,rainbow_bunch7}. Here, dimer are established  among symmetric qubits with respect to the center: $i_k = k, j_k = N +1-k.$ The state exhibits volume-law scaling of entanglement entropy with the increase of the system size. In this case also, all the single-site entropies become $S(\rho_i)=\log(2).$ Whereas, the non-zero values of the entanglement entropies can be obtained only for $\rho_{ij}$ such that $i+j\neq N+1$, given by $S(\rho_{ij})=2\log(2)$. As a result, we get $J_{ij}=\log(2)$, for $i+j=N+1$ and zero otherwise. Interestingly,  one can note that in this case, the entanglement adjacency matrix is not emerging as a restriction on the adjacency matrix representing the Hamiltonian. In other words, an observer trying to determine the geometry from the distribution of the entanglement will not find the correct geometry of the Hamiltonian.\

{\it {\bf III.} GHZ state.-} A different case  we consider here is the $N$-party GHZ state, expressed as $|GHZ\rangle=\frac{1}{\sqrt{2}}(|0\rangle ^N|+|1\rangle ^N)$. In this case, the entropy values of all the bipartitions, irrespective of the number  of sites, become identical, given by $\log(2)$. Hence,  all the $J_{ij}$'s become same. As a result,   to represent the block entropies using our formalism, we consider $J_{ij}=0  \hspace{.1cm} \forall   i, j \in N$ and put the value of the constant term in Eq. (\ref{entropy_general}), $s_0=\log(2)$. This suggests that the GHZ state does not have a geometrical interpretation in this framework.


\subsection{Numerical computation} 
In this section, we describe the numerical methodology to obtain the entanglement adjacency matrix for which Eq. (\ref{entropy_general}) is not exact. 
For any general block,  the  relation between
parameters and entropies can be expressed through
\begin{eqnarray}
\sum_{(ij)}{\mathcal D}_{I,(ij)} J_{ij} = S_I,
\label{eq:defA}
\end{eqnarray}
where $I=(x_1\cdots x_N)$ denotes the binary expansion for the index
of each block, i.e. $x_k=1$ if site $k$ belongs to block $I$ (and zero
otherwise), and ${\cal D}_{(x_1\cdots x_N),(ij)}=1$ 
if $(x_i,  x_j) =(0,1)$ or (1,0),  and zero otherwise. In our case, vector $J$ contains all the $J_{ij}$ in order, i.e., has dimension $N(N - 1)/2$, while vector $S_I$ contains all the entropies, so it has dimension $2^{N}$. Thus, matrix $\mathcal{D}$ has dimension $2^N \times N(N -1)/2$. In other terms: as many rows as entropies, and  as many columns as couplings.  \

Eq. \eqref{eq:defA} is a strongly overdetermined linear
system which will be, in general, incompatible. Yet, it is
possible to find an approximate solution in the least-squares
sense, using the equation 
\begin{eqnarray}
\sum_{(i'j')}  {\cal D}^\dagger {\cal D}_{(ij),(i'j')} J_{i'j'}=
  \sum_I {\cal D}_{I,(ij)} S_I.
\label{eq:normaleqs}
\end{eqnarray}
Subsequently, an estimation of  the relative error made in this optimization process can be made as follows. 
Let $\hat S_I$ be the estimate obtained through Eq. (\ref{eq:normaleqs}). The
error will be defined as
\begin{eqnarray}
   {\cal E}={1\over 2^N} \sum_{I=0}^{2^N-1} \left| S_I- \hat S_I \right|.
\label{eq:error}
\end{eqnarray}
In the forthcoming section, we will use this formula to compute the error made in computation of  entanglement adjacency matrix for various physical models.


\section{Entanglement contour} 

In this section, we discuss another important facet of our formalism, where a more refined approach to characterize entanglement entropy of any bipartition is presented in terms of the entanglement contour function introduced earlier in the literature \cite{contour1} (see also \cite{rainbow_bunch4,contour2,contour3,contour4,contour5,contour6,contour7}). The entanglement contour function for a given block estimates the contribution of each site to the entanglement entropy  obtained for that bipartition. For a given block $A$, mathematically it can be expressed as
\beq S_A =\sum_{i \in A} s_A(i), \qquad s_A(i) \geq 0\,.
\label{eq:def_contour}\eeq
Interestingly, from Eq. (\ref{entropy_general}) one can observe that the entanglement adjacency matrix provides a natural entanglement contour which can be expressed as
\beq\quad s_A(i)\equiv \sum_{j\in A^c} J_{ij}  \, .
\label{eq:contour_J}
\eeq
The contour function defined above satisfies all the standard constraints listed in Ref. \cite{contour1}. Here, we stress the fact that unlike the actual formulation of the entanglement contour introduced in Ref. \cite{contour1}, our approach aims to provide an overall description of bipartite entanglement by considering contributions of {\it all} bipartitions and not just the ones consisting of simply connected intervals. Moreover, the formalism includes any general quantum systems, including interacting cases.

\begin{figure}[h]
\begin{center}
  \includegraphics[width=10cm]{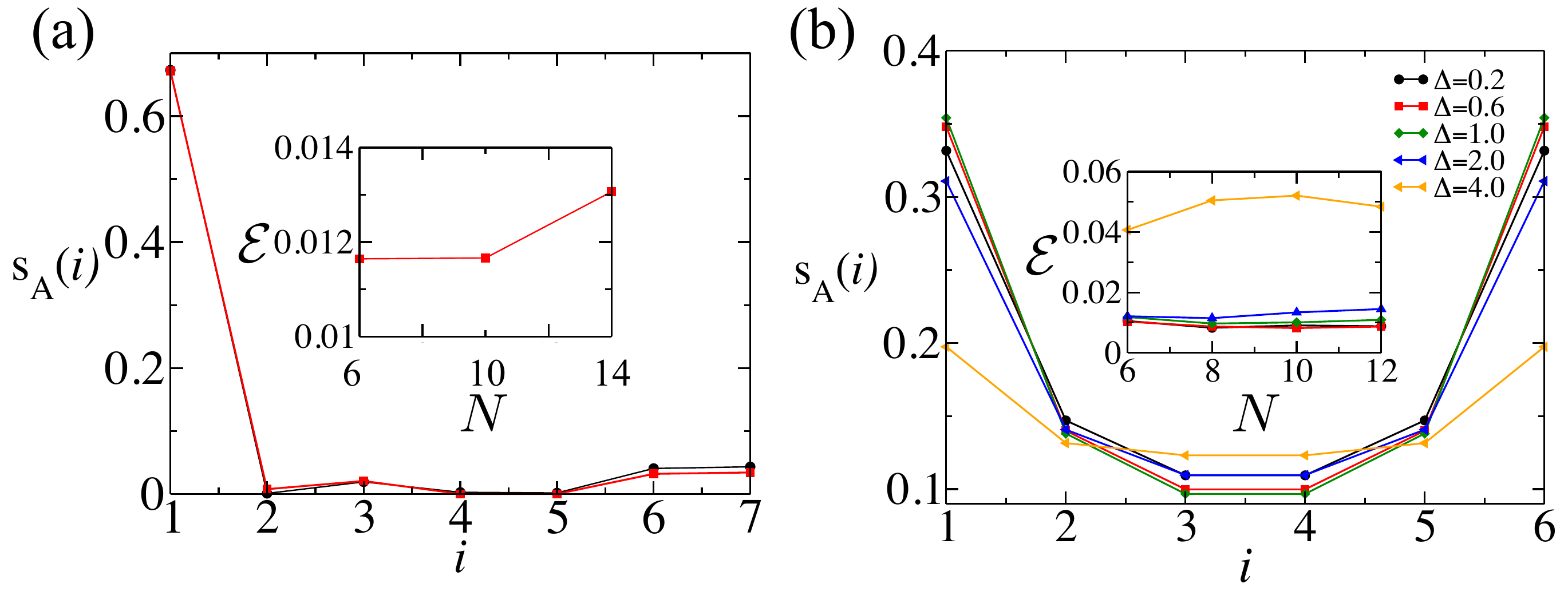}
  \caption{In panel (a), we compare the  contour functions for the entanglement entropy $s_A(i)$, those obtained using the Eqs.  (\ref{eq:contour_J}) and (\ref{eq:vidal_contour}),    for the  ground state of dimerized Hamiltonian ($t_{ij} = (1+\delta (-1)^i), |i-j| = 1,\delta=0.5$, $N=14$).   Whereas, in panel (b), we plot the contour functions for the entanglement entropy obtained  using Eq.  (\ref{eq:contour_J}) for ground state of  $XXZ$ Hamiltonian for different values of the parameter $\Delta$, and for $N=12$. Additionally, in the inset of both the figures, the scaling of the error ($\mathcal{E}$) with the system size ($N$) have been shown for all the  parameter values considered.}
  \label{contour_all}
  \end{center}
  \end{figure}

 {\it Contour plot for free-fermionic Hamiltonian:} For free-fermionic model, described below 
\begin{eqnarray}
\mathcal{H}_{free-ferm}=-\frac{1}{2}\sum_{ij}t_{ij}(c^{\dagger}_i c_{j}+hc),
\label{Ham_free_ferm}
\end{eqnarray}
where $c_i (c^{\dagger}_i)$'s is the fermioninc anhillation (creation) operators at site $i$, and $t_{ij}$ is the  hopping matrix, a proposal for the contour is given in Ref. \cite{contour1}, 
\beq
s_A(i)= \sum_{p=1}^{|A|} |\Phi_{p,i}^{(A)} |^2  \;  H(\nu_p),
\label{eq:vidal_contour}
\eeq
where $\Phi_{p,i}^{(A)}$ is the eigenvector with eigenvalue $\nu_p$,  of the correlation matrix block\cite{rainbow_bunch0,fermi_red_den} restricted to  $A$ and $H(x) = -  \left[ x \log x + (1- x)\log (1-x) \right] $. Using the above equation, in Fig.  \ref{contour_all}(a), we compute the entanglement contour function for the ground state of the dimerized Hamiltonian, which can be obtained from the  free-fermionic model described in Eq. (\ref{Ham_free_ferm}), for $t_{ij} = (1+\delta (-1)^i), |i-j| = 1$ and compare that to the contour function obtained using the elements of the entanglement adjacency matrix,  $J_{ij}$, as described in Eq. (\ref{eq:contour_J}). From the figure, we note that the contour functions obtained using these two different methods are very similar to each other.

   {\it Contour plot for $XXZ$ Hamiltonian:} Subsequently,  we move one step further and apply the formalism to an interacting model, namely, the one-dimensional $XXZ$ model, expressed as
   \beq
H_{XXZ}=\sum_i^N S_i^{x}S_{i+1}^{x}+S_i^{y}S_{i+1}^{y}+\Delta S_i^{z}S_{i+1}^{z},
\label{eqn:xxz}
\eeq
   where $S_l^k (k\in x,y,z)$ are the Pauli operators at site $l$, and $\Delta$ is the anisotropy along the $z$-direction.  Note that in this case, to obtain the set of entropies for all possible bipartitions of the ground state of the model, we perform  the exact diagonalization method. The behavior of the entanglement contour function obtained for the half-chain, for the critical ($\Delta \leq  1$) and non-critical ($\Delta>1$) cases are depicted in Fig.        \ref{contour_all}(b). 
   

 \section{Entanglement current}  
In this section, we extend our formalism to  one-dimensional conformal invariant systems and  attempt to provide an interpretation of the entanglement adjacency matrix entries,  $J_{ij}$,  as the two-point correlator of an entanglement current operator.   The entanglement entropy of the ground state of a CFT for an interval $A = (u, v)$ embedded  in the infinite line is given by
\beq
S_A = \frac{c}{3} \log \frac{ v - u}{\epsilon}, \
 \label{cft1_text}
\eeq
where $c$ is the central charge and $\epsilon > 0$  a UV  cut-off. One can note that Eq. (\ref{cft1_text})  can be obtained from a continuous version of Eq.  (\ref{entropy_general})\beq
S_A =   \frac{c}{6}  \int_{A_\epsilon} dx \int_{ A^c} dy \; J(x,y)  \, ,
\label{cft2_text}
\eeq
with  $A_\epsilon =( u+ \epsilon, v - \epsilon)$ and $A^c = (- \infty, u) \cup (v, \infty)$, by  choosing
\beq
J(x,y)  = \frac{1}{(x-y)^2}  \, .
\label{cft3_text}
\eeq
This equation indicates  that $J(x,y)$ is the two point correlator, on the complex plane,  of a current operator ${\bf J}$, whose integration along segments, as in Eq. (\ref{cft2_text}),  is invariant under reparametrization. $J(x,y) dx dy$ represents the  amount of entanglement between the intervals $(x, x+dx)$ and $(y, y+dy)$. This field theory realization leads to think of entanglement as a flow among the parts of the system, in analogy to the flow of energy that is characterized by the stress tensor. Moreover, using the construction described in Ref. \cite{cft4}  for entanglement Hamiltonians in CFT, one can show that Eq. (\ref{cft2_text}) reproduces the values of $S_A$, for the space-time geometries $\Sigma$, that are conformally equivalent  to an annulus. In these cases $J(x,y)$  is given by the two point correlator $
J(x,y)  = \langle {\bf J}(x) \,  {\bf J}(y) \rangle_\Sigma.$
Notice that in the conformal field theory systems the representation is exact only when $A$ is an  interval, but not in general.


\section{Conclusions}
To conclude, in this work, we aimed to unveil the geometry revealed from the entanglement properties of any pure quantum state through the elements of a generalized adjacency matrix, such that the entropy values of any bipartition of the state can be approximated as a weighted sum of all the links connecting the sites across that bipartition. We reported certain examples, where the optimal geometry emerged from the entanglement structure, turned out to be completely different from that suggested by the parent Hamiltonian of the system.  Subsequently, we showed that  our formalism provided a natural route to compute the entanglement contour, introduced earlier for the non-interacting models, which essentially helped us to extend the concept for interacting models as well.  Finally, we showed that for conformal invariant systems, a more insightful explanation of the elements of such generalized adjacency matrices can be obtained in terms of a two-point correlator of an entanglement current operator.\\

{\bf Acknowledgements.} GS would like to thank William Witczak-Krempa for the invitation to participate in the  Quantum Theory and Symmetry XI conference held in Montreal in July 2019.  We acknowledge financial support from the grants  PGC2018-095862-B-C21, QUITEMAD+ S2013/ICE-2801,  SEV-2016-0597 of the ``Centro de Excelencia Severo Ochoa” Programme and the CSIC Research Platform on Quantum Technologies PTI-001.

%

\end{document}